\documentstyle[aps,multicol,psfig]{revtex}

\begin{document}

\title{Nuclear  Pairing  in the T=0 channel revisited}
\author{E. Garrido$^{1}$, P. Sarriguren$^{1}$, E. Moya de Guerra$^{1}$, U. 
Lombardo$^{2}$,  P. Schuck$^{3}$, \\ 
H.J. Schulze$^{4}$}
\address{$^{1}$ Instituto de Estructura de la Materia, Consejo Superior de 
Investigaciones Cient\'{\i }ficas, \\
Serrano 123, E-28006 Madrid, Spain\\
$^{2}$ Dipartimento di Fisica, Universit\`a di Catania,57 Corso Italia, 
I-95129 Catania, Italy \\
$^{3}$ Institut de Physique Nucl\'{e}aire, Universit\'{e} Paris-Sud,
F-91406 Orsay Cedex, France\\
$^{4}$ Departament d'Estructura i Constituents de la Mat\`{e}ria, Universitat 
de Barcelona\\
Av. Diagonal 647, E-08028 Barcelona, Spain}
\date{\today }

\maketitle

\begin{abstract}
Recent published data on the isoscalar gap in symmetric nuclear matter using
the Paris force and the corresponding BHF single particle dispersion are
corrected leading to an extremely high proton-neutron gap of $\Delta \sim 8$
MeV at $\rho \sim 0.5\rho_0$. Arguments whether this value can be reduced
due to screening effects are discussed. A density dependent delta interaction
with cut off is adjusted so as to approximately reproduce the nuclear matter
values with the Paris force.
\end{abstract}


\widetext
\tighten

\begin{multicols}{2}

In a recent publication \cite{gar99} the possibility to reproduce the gap in 
nuclear matter, as obtained e.g. from the Paris NN force, by an effective 
density dependent zero range force, was investigated. Supplied with 
an energy cut off such effective forces turned indeed out to be able to
reproduce very reasonably the gap values in the isospin T=0 and T=1 channels 
over the whole relevant range of densities. The adjustments were performed
on previously published solutions of the gap equation using 
Brueckner-Hartree-Fock results for the single particle spectra \cite{bls}.
Such effective forces may possess some analogies with similar ones
frequently used in recent structure calculation of superfluid nuclei 
\cite{terasaki}. Unfortunately, due to the subtleties connected with the 
numerical solution of the gap equation, the published results in the T=0 
channel were not accurate enough so that the corresponding gap is 
underestimated in \cite{gar99,bls} by about 20\%. It is the purpose of this 
note to give the corrected results for the gap in the T=0 channel and also 
to readjust the corresponding density-dependent $\delta$-force. We also 
discuss again the issue whether screening affects the T=1 and T=0 channels 
differently.

In Fig. 1 we show the correct result for the isoscalar gap as obtained with 
the Paris force \cite{paris} using two independent numerical codes. We also 
checked that the Argonne V14 force \cite{arg} gives practically the same result.
What is striking is the giant gap value of $\sim 8$ MeV at maximum, which is 
of the same order as the Fermi energy at the corresponding density. Even 
around saturation, $\Delta $ is still of the order of several MeV. 
This is clearly a strong coupling situation as expected from the fact that at
low density the n-p Cooper pair turns into the deuteron wave 
function \cite{bls}. The above values are actually much more compatible with 
earlier calculations of the critical temperature in Ref. \cite{alm} than the 
previous results \cite{bls}. Indeed, considering the usual relation 
$\Delta =1.76\ T_c$ \cite{fetter}, quantitative agreement between
the results of \cite{alm} and the ones in Fig. 1 is obtained. 
In order to obtain an estimate of the typical magnitude of the isoscalar gap 
in a finite nucleus, we apply the local density approximation and average 
the local gap over the density at the Fermi energy. This procedure has given 
reliable estimates of the average energy dependent gap in the isovector 
channel \cite{sch}. We therefore calculate

\begin{figure}
\centerline{\psfig{figure=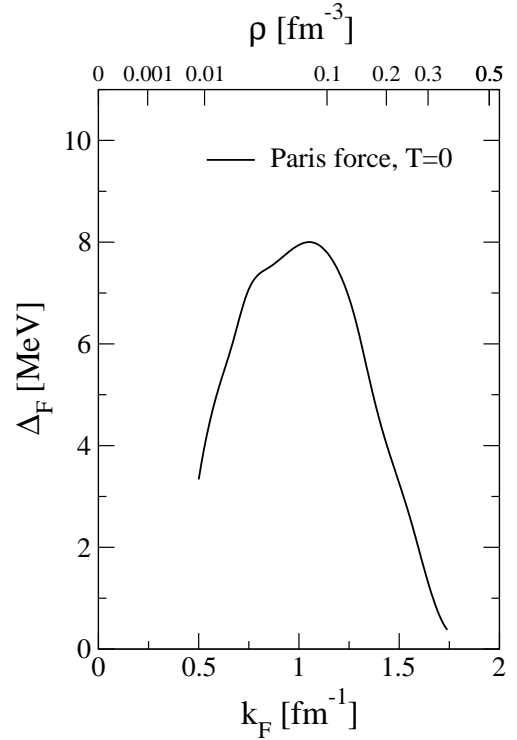,width=9.5cm,%
bbllx=1.5cm,bblly=3.6cm,bburx=20.2cm,bbury=24.6cm}}
\vspace*{0.3cm}
\caption[]{Pairing gap versus Fermi momentum for symmetric nuclear matter
in the T=0 channel from the Paris potential.}
\end{figure}

\begin{equation}
\Delta =\frac{\int dr r^2 \Delta (k_F(r))k_F(r)}{\int dr r^2 k_F(r)} \ ,
\end{equation}
where the local Fermi momentum is defined as

\begin{equation}
k_F(r)=\sqrt{(\mu-V(r)) 2m/\hbar^2} \ ,
\end{equation}
with $\mu $ the chemical potential. We take the same single particle 
potential $V(r)$ as in \cite{sch} and the result for e.g. N=Z=35 is that 
$\Delta$ is of the order of 3 MeV. Compared to the neutron-neutron and 
proton-proton channels this is a very high value. 

We already discussed in \cite{gar99} and show again in Fig. 2 that the use 
of the Paris force in conjunction with the k-mass, $m^*/m$, yields gap 
values as a function of density which are globally very similar to the 
ones of the Gogny force for T=1 and therefore, the use of a bare force 
seems not unreasonable in the T=1 channel. The fact that $\Delta$ for 
T=1 drops off quite a bit faster close to saturation for the Paris force 
than for the Gogny D1S force may be attenuated in a finite nucleus to 
quite some extent, since a certain averaging over all densities 
$\rho < \rho_0$ takes place. Therefore the needed medium renormalization 
of the bare force seems to be of minor importance in the T=1 
channel\footnote{Of course it cannot be excluded that the medium completely
re-shuffles the distribution of gap values, still reproducing experimental
pairing phenomena in finite nuclei}.
However, the situation may not be the same for T=0 pairing. The extremely 
strong T=0 pairing stems essentially from the fact that with respect to 
the T=1 channel the tensor force is acting additionally. Without the 
tensor force $np$ (T=0) and $nn$ (T=1) pairing would be of comparable 
magnitude. The screening of the tensor force in the medium is, however, 
still a controversial subject \cite{zheng}. On the other hand, even for 
very low densities where screening should not be so important, T=0 pairing 
remains strong. Therefore, there may be a good chance that the new heavier 
exotic nuclei with $N=Z$ experience quite pronounced $np$ superfluidity. 
This may well be the cause for the so called Wigner energy of the nuclear 
mass formula, since it can be shown \cite{wigner} that away from symmetric 
nuclei, T=0 pairing looses very quickly its strength.

Let us now proceed to the readjustment of the effective T=0 delta force. We
use the standard ansatz \cite{gar99,esb}

\begin{eqnarray}
v(\vec{r_1},\vec{r_2})&=&v_0\left\{ 1-\eta \left[ \rho \left( \frac{r_1+r_2}{2}
\right)/\rho_0\right] ^\alpha \right\} \times \nonumber \\ & & 
\delta (\vec{r_1}-\vec{r_2}) 
\left( 1+P_\sigma \right)/2 \ .
\label{eqv}
\end{eqnarray}
With the above density-dependent zero range force, the gap equation reads

\begin{figure}
\centerline{\psfig{figure=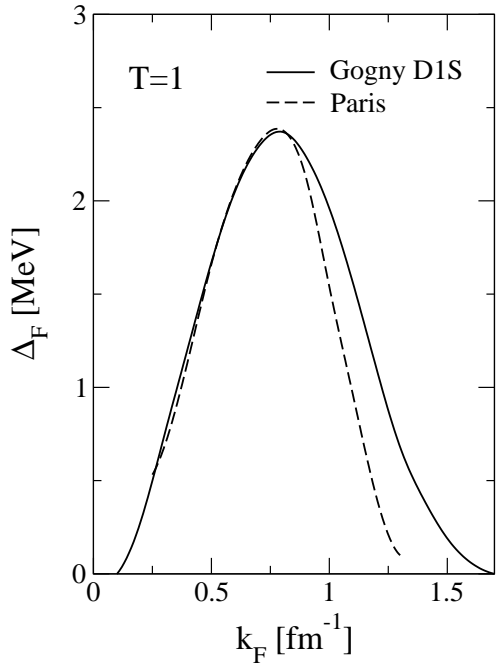,width=9.5cm,%
bbllx=1.5cm,bblly=0.6cm,bburx=20.2cm,bbury=21.6cm}}
\vspace*{-1.0cm}
\caption[]{Pairing gap $\Delta_F$ in the $^1S_0$ channel in symmetric
nuclear matter calculated with the Gogny force D1S compared with results
from the Paris force.} 
\end{figure}

\begin{eqnarray}
1 &= &
-\frac{v_{0}}{\pi ^{2}}\left[ 1-\eta \left( \rho /\rho _{0}\right)
^{\alpha }\right] \left( \frac{m^{*}\left( \rho \right) }
{2\hbar ^{2}}\right) ^{3/2} \times \nonumber \\ && 
\int_{0}^{\epsilon _{C}}d\epsilon 
\sqrt{\frac{\epsilon }{\left( \epsilon -\epsilon _{F}\right) ^{2}
+\Delta ^{2}}} \ . \label{eqber}
\end{eqnarray}

In Fig. 3 we present two fits for the above ansatz, one of the fits is 
obtained from the following parameters: 
$\alpha =0.2$, $\eta=-0.10$ and a cut off energy $\epsilon _C=60$ MeV 
(see Ref.\cite{gar99}), using the effective mass $m^*/m$ as obtained 
from the Gogny force. The other fit is obtained  by using a bare mass 
and parameters $\alpha =0.90$, $\eta =0.40$ and $\epsilon _C=60$ MeV. 

As one can see in Fig. 3, the fit obtained using the bare mass is able to
reproduce the microscopic calculation up to the highest values of $k_F$
($k_F\sim 1.7$ fm$^{-1}$), while the fit obtained using the effective mass
breaks down at lower densities corresponding to $k_F\sim 1.35$ fm$^{-1}$.
The reason for this different behavior can be traced back to the dependence
on the effective mass inside the integral of the gap equation. It turns out 
that in order to get a solution of the gap equation (\ref{eqber}), the 
energy cut off $\epsilon_C$ should be larger than the Fermi energy 
$\epsilon_F$. Otherwise no value of $\Delta$ satisfies the equation. 
In the case of the energy cut off used in Figs. 3 and 4 ($\epsilon_C=60$ MeV), 
the largest $k_F$ reachable is $k_F\sim 1.7$ fm$^{-1}$ when bare masses are 
used, but only  $k_F\sim 1.35$ when effective masses are used instead. 
Therefore, we plot in Figs. 3 and 4, the fits obtained only up to those 
values of $k_F$, when $m^*/m\neq 1$. Nevertheless, the fits cover all the 
physically relevant range of densities form zero to saturation
($\rho_0=0.16$ fm$^{-3}$). 

\begin{figure}
\centerline{\psfig{figure=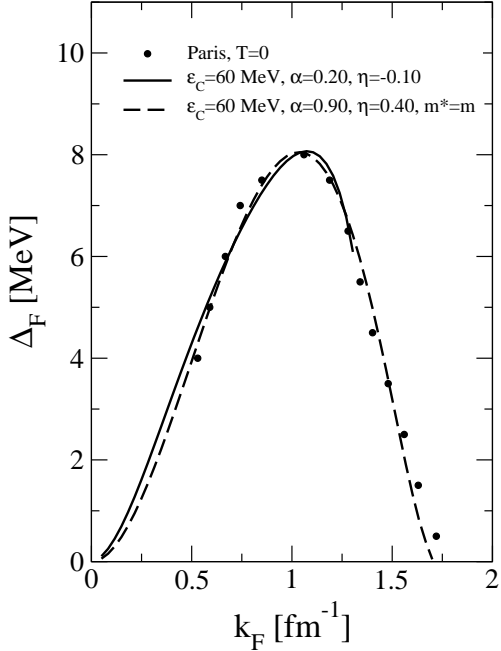,width=9.5cm,%
bbllx=1.5cm,bblly=0.6cm,bburx=20.2cm,bbury=21.6cm}}
\vspace*{-1.0cm}
\caption[]{T=0 pairing gap in nuclear matter. The dots are the results 
obtained for the Paris potential. The curves are fits with Eq.(\ref{eqv}) 
using  an energy cut off $\epsilon_C=60$ MeV, $v_0= -480$ MeV fm$^3$, and 
different parameters for the fit with effective mass $m^*$ (solid line, 
$\eta=-0.10,\; \alpha=0.20$) and for the fit with the bare mass (dashed 
line, $\eta=0.40,\; \alpha=0.90$).}
\end{figure}

In principle, in the T=0 channel, $v_0$ should be chosen such that the
deuteron binding energy is reproduced in free space. However, we have
found that with this condition the fit obtained is very poor. Therefore,
for a given energy cut $\epsilon_C$, we vary the parameter $v_0$ from
the value that produces a bound state at zero energy 
$v_0=-(\hbar ^2/m)(2\pi^2/\sqrt{2m\epsilon_C})$, 
up to the value that produces the bound state at the deuteron energy \cite{esb},
and choose the best fit. The fits in Fig. 3 have been obtained with 
$v_0=-480$ MeV fm$^3$ as it corresponds to a bound state at zero energy. 
This reduces the value of the gap at low densities but improves significantly 
the fit at higher energies. On the other hand, as we shall see in Fig. 4, 
the value  of $v_0$ is chosen between the two extreme values considered, 
bound state at zero energy and at the deuteron energy. 
In any case, the values used for $v_0$ are quoted in each case.

\begin{figure}
\centerline{\psfig{figure=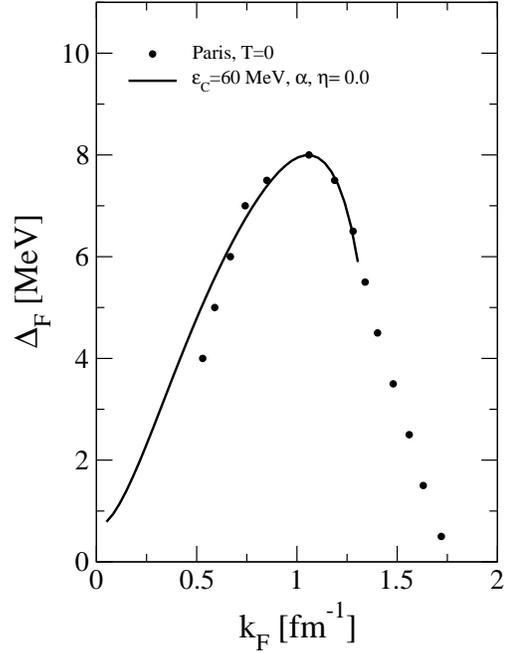,width=9.5cm,%
bbllx=1.5cm,bblly=0.6cm,bburx=20.2cm,bbury=21.6cm}}
\vspace*{-1.0cm}
\caption[]{Same as in Fig. 3 but suppressing the density dependence
($\eta=0$) and using $v_0= -530$ MeV fm$^3$.}
\end{figure}

In Fig. 4 we present a similar fit for the case with $m^*/m\neq 1$, however, 
suppressing the density dependence completely, that is $\eta =0$. Since this 
parameter was already small for the case in Fig. 3 the fit is still acceptable 
and only a slight deterioration  at the low density end is visible. 
Let us mention also that the use of the bare mass $m^*=m$ allows an excellent 
fit of the microscopically calculated gap values at all densities (see Fig. 3). 
However, realistic calculations of finite nuclei are rarely performed with 
the bare nucleon mass.

As a first guess we may try to use the effective pairing force obtained with 
the present  fit also for finite nucleus calculations. This will give a rough 
account of whether the use of a bare force in a finite nucleus is at all 
reasonable in the T=0 channel.
We would, however, like to point out that the expression of Eq.(\ref{eqv}) 
for finite nuclei may not give precise reproduction of the results one 
would obtain with a direct use of the Paris force in the gap equation.
Indeed, in the mean time, we compared in the T=1 channel the results of
the genuine Gogny force and its density dependent $\delta$-force substitute
elaborated in \cite{gar99} in a half infinite matter calculation \cite{farine}.
Preliminary results show that the detailed surface dependence of the gap
and of the anomalous density seem to be quite different in both cases. 
However, integrated quantities like the correlation energy may still be rather 
similar.

Of course, it should be interesting for the future to derive also an effective 
finite range force in the T=0 channel which is as efficient as the Gogny force 
for T=1 pairing. In fact the Gogny force has never been used for $np$ pairing. 
However, since in this channel the density dependent zero range force enters, 
one has to introduce an additional cut off which is an unknown adjustable 
parameter.

In summary we give corrected values of the $np$ (T=0) gap in nuclear matter
using the Paris force together with Brueckner-Hartree-Fock single-particle
energies. An extremely high value of $\Delta \sim 8$ MeV at 
$\rho\sim 0.5\rho_0$ is obtained, leading
to a gap value in finite nuclei of $\sim$ 3 MeV. Arguments are
advanced that the pairing force in the T=0 channel  may be more strongly
screened than in the T=1 channel. We then adjust a density-dependent 
$\delta-$force to the nuclear matter gap values. The fit is reasonably
successful for densities below saturation.

\acknowledgements

This work was supported in part by DGESIC (Spain) under Contract number 
PB98-0676, by the {\it Groupement de Recherche: Noyaux Exotiques}, CNRS-IN2P3,
and by the programs {\it Estancias de cient\'\i ficos y tecn\'ologos
extranjeros en Espa\~na}, SGR98-11 (Generalitat de Catalunya), and DGICYT 
(Spain) no. PB98-1247.

\samepage

\end{multicols}

\end{document}